\documentstyle [12pt]{article}
\begin{document}

\title{On correspondence between tensors and bispinors}

\author{M.V. Gorbatenko, A.V. Pushkin }
\maketitle

\begin{center}
{RFNC-VNIIEF, Sarov, Nizhni Novgorod Reg., Russia, 607190\\E-mail:
gorbatenko@vniief.ru; pav@vniief.ru  }
\end{center}

\vskip5mm

1. It is known that in the four-dimensional Riemannian space the complex 
bispinor generates a number of tensors: scalar, pseudo-scalar, vector, 
pseudo-vector, antisymmetric tensor. This paper solves the inverse problem: 
the above tensors are arbitrarily given, it is necessary to find a bispinor 
(bispinors) reproducing the tensors. The algorithm for this mapping 
constitutes construction of Hermitean matrix $M$ from the tensors and 
finding its eigenvalue spectrum. A solution to the inverse problem exists 
only when $M$ is nonnegatively definite. Under this condition a matrix 
$Z$satisfying equation $M = ZZ^{ +} $ can be found. One and the same system 
of tensor values can be used to construct the matrix $Z$ accurate to an 
arbitrary factor on the left-hand side, viz. unitary matrix $U$ in polar 
expansion $Z = H \cdot U$. The matrix $Z$ is shown to be expandable to a set 
of bispinors, for which the unitary matrix $U$ is responsible for the 
internal (gauge) degrees of freedom. Thus, a group of gauge transformations 
depends only on the Riemannian space dimension, signature, and the number 
field used. 
The constructed algorithm for mapping tensors to bispinors admits extension 
to Riemannian spaces of a higher dimension.

2. Bispinor matrix$Z$\footnote{%
The term ``paper'' means a reference to two papers by M.V.Gorbatenko and
A.V.Pushkin ``On correspondence between tensors and bispinors'' [1], [2].}

At a given field of metric tensor, $g_{\alpha \beta} \left( {x} \right)$,
the field of Dirac matrices (DM), $\gamma _{\alpha} \left( {x} \right)$, is
determined by relation
\begin{equation}  \label{eq1}
\left[ {\gamma _{\alpha} \left( {x} \right)\;,\;\gamma _{\beta} \left( {x}
\right)} \right]_{ +} = 2g_{\alpha \beta} \left( {x} \right) \quad .
\end{equation}
Hereafter we call DM $\gamma _{\alpha} \left( {x} \right)$ as world matrices.

Suppose, the DM are implemented over the real number field. In this case, if
relation (\ref{eq1}) is satisfied by DM $\gamma _{\alpha} \left( {x} \right)$%
, it is satisfied by three more DM systems, namely: $- \gamma _{\alpha}
\left( {x} \right);\;\gamma _{\alpha} ^{T} \left( {x} \right);\; - \gamma
_{\alpha }^{T} \left( {x} \right)$. Here $\gamma _{\alpha} ^{T} \left( {x}
\right)$ are DM produced from DM $\gamma _{\alpha} \left( {x} \right)$
through transposition.

The existing discrete transposition operation allows construction of
nontrivial fields of matrix scalars, vectors, and other tensors from DM. A
simplest example of the matrix scalar is $\left( {\gamma ^{\nu} \left( {x}
\right)\gamma _{\nu} ^{T} \left( {x} \right)} \right)$. It is easy to notice
that at DM transformations by rule
\begin{equation}  \label{eq2}
\gamma _{\alpha} \left( {x} \right) \to {\gamma} ^{\prime}_{\alpha} \left( {x%
} \right) = T\left( {x} \right)\gamma _{\alpha} \left( {x} \right)T^{ -
1}\left( {x} \right),
\end{equation}
DM $- \gamma _{\alpha} \left( {x} \right);\;\gamma _{\alpha} ^{T} \left( {x}
\right);\; - \gamma _{\alpha} ^{T} \left( {x} \right)$are transformed
similarly, provided the nonsingular matrices $T\left( {x} \right)$ are
orthogonal, that is
\begin{equation}  \label{eq3}
T^{T}\left( {x} \right) = T^{ - 1}\left( {x} \right).
\end{equation}
Hence, any
relations involving matrix tensors, which have been constructed involving $%
\gamma _{\alpha} \left( {x} \right)$; $- \gamma _{\alpha} \left( {x}
\right);\;\gamma _{\alpha} ^{T} \left( {x} \right);\; - \gamma _{\alpha
}^{T} \left( {x} \right)$, retain their form at transformations (\ref{eq3})
with orthogonal matrices $T\left( {x} \right)$. We treat this feature of
matrix relations as $T\left( {x} \right)$ invariance of the DM apparatus. At
each Riemannian space point a set of orthogonal matrices form group $O\left( 
{4,R} \right)$. $T\left( {x} \right)$ invariance of the DM apparatus can be
interpreted as an invariance with respect to a manner of matrix row and
column numbering.

Introduce a field of frame-of-reference vectors $H_{\alpha} ^{k} \left( {x}
\right)$; these satisfy relations\footnote{%
Greek and Latin letters take on the same values, $0,\,1,\,2,\,3$. The
difference in the letters is that the Greek letters mean that the value
responds to transformations of the world coordinates in the Riemannian
space, while the Roman ones imply responding to coordinate transformations
in the local tangent space. The Roman subscripts from the alphabet beginning
also denote space subscripts $\,1,\,2,\,3$ hereinafter.}
\begin{equation}  \label{eq4}
g_{\alpha \beta} \left( {x} \right) = H_{\alpha} ^{m} \left( {x}
\right)H_{\beta} ^{n} \left( {x} \right)g_{mn} \quad ,
\end{equation}
where $g_{mn} $ is a metric tensor in the Minkowski tangent space
at the points having coordinates $\left( {x} \right)$. Assume that Galilean
coordinates are used in the tangent spaces, so that tensor $g_{mn} $ has
diagonal form, with numbers $\left( {\ - 1,\;1,\;1,\;1} \right)$ appearing
along the diagonal. At the world coordinate transformation $x \to {x}%
^{\prime}\left( {x} \right) \quad H_{\alpha} ^{k} \left( {x} \right)$ are
transformed by the law of ordinary vectors,
\begin{equation}  \label{eq5}
H_{\alpha} ^{k} \left( {x} \right) \to {H^{\prime}}_{\alpha} ^{k} \left( {{x}%
^{\prime}} \right) = \left( {{{\partial x^{\mu} } \mathord{\left/ {\vphantom
{{\partial x^{\mu} } {\partial {x}'^{\alpha} }}} \right.
\kern-\nulldelimiterspace} {\partial {x^{\prime}}^{\alpha} }}} \right) \cdot
H_{\mu} ^{k} \left( {x} \right).
\end{equation}
$H_{\alpha} ^{k} \left( {x} \right)$ values also respond to transformations
of coordinates in the tangent spaces. Under our assumptions on the choice of
the Galilean coordinates in the tangent spaces, the tangent space coordinate
transformations are Lorentz transformations,
\begin{equation}
H_{\alpha }^{k}\left( {x}\right) \to {H}_{\alpha }^{\prime }{}^{k}\left( {x}%
\right) =w^{k}{}_{p}\left( {x}\right) \cdot H_{\alpha }^{p}\left( {x}\right)
.  \label{eq6}
\end{equation}
$w^{k}{}_{p}\left( {x}\right) $ satisfy the relations
\begin{equation}
w^{m}{}_{p}\left( {x}\right) w^{n}{}_{q}\left( {x}\right) \cdot
g_{mn}=g_{pq}.  \label{eq7}
\end{equation}
Relations (\ref{eq7}) mean in essence that the metric $g_{mn} $ remains
unchained at the Lorentz transformations.

The availability of frame-of-reference vectors $H_{\alpha} ^{k} \left( {x}
\right)$ at each Riemannian space point allows introduction, alongside the
world Dirac matrices $\gamma _{\alpha} \left( {x} \right)$, one more DM
type, i.e. local DM $\gamma _{k} \left( {x} \right)$,
\begin{equation}  \label{eq8}
\gamma _{k} \left( {x} \right) = H_{k}^{\alpha} \left( {x} \right)\gamma
_{\alpha} \left( {x} \right).
\end{equation}
The local DM satisfy the relation
\begin{equation}  \label{eq9}
\left[ {\gamma _{m} \left( {x} \right)\;,\;\gamma _{n} \left( {x} \right)}
\right]_{ +} = 2g_{mn} \quad .
\end{equation}

A feature of relation (\ref{eq9}) is that its right-hand side is independent
on coordinates, while $\gamma _{k} \left( {x} \right)$ appearing in the
left-hand side are, generally speaking, dependent on them. At DM world
coordinate transformations $\gamma _{k} \left( {x} \right)$ behave like
scalars, while at Lorentz transformations $\gamma _{k} \left( {x} \right)$
behave similar to frame-of-reference vectors. As a result, relation (\ref
{eq8}) retains its form at either transformation type. At $T\left( {x}
\right)$ transformations of (\ref{eq3}) we have:
\begin{equation}  \label{eq10}
\gamma _{k} \left( {x} \right) \to {\gamma} ^{\prime}_{k} \left( {x} \right)
= T\left( {x} \right)\gamma _{k} \left( {x} \right)T^{ - 1}\left( {x}
\right),
\end{equation}
so that (\ref{eq9}) form is unchanged.

At each Riemannian space point, alongside the world DM, $\gamma _{\alpha}
\left( {x} \right)$, and the local DM, $\gamma _{k} \left( {x} \right)$, one
more DM system can be introduced without loss of generality, which we call
as ``doubly local DM system'' and denote as $\mathop {\gamma} \limits^{0}
_{k} $. Formally, DM $\mathop {\gamma} \limits^{0} _{k} $, like $\gamma _{k}
\left( {x} \right)$, are introduced using a relation similar to (\ref{eq9}),
that is relation
\begin{equation}
[ {\mathop {\gamma }\limits^{0}{}_{m},\;\mathop {\gamma }%
\limits^{0}{}_{n}}] _{+}=2g_{mn}.  \label{eq11}
\end{equation}
However $\mathop {\gamma} \limits^{0} {}_{k} $ differ from $\gamma _{k} \left( 
{x} \right)$ in two features. First, these are independent on coordinates.
Second, these do not change at Lorentz transformations in tangent spaces.
The second difference can be valid only under the condition that the matrix
subscripts in $\mathop {\gamma} \limits^{0} _{k} $ are of another nature
than those in $\gamma _{k} \left( {x} \right)$. The difference in the matrix
subscripts manifests itself at Lorentz transformations:
\begin{equation}
\left\{ {\ 
\begin{array}{l}
{\gamma _{k}\left( {x}\right) \to {\gamma }_{k}^{\prime }\left( {x}\right)
=w_{k}^{p}\left( {x}\right) \gamma _{p}\left( {x}\right) \;,} \\ 
{\mathop {\gamma }\limits^{0}{}_{k}\to \mathop {{\gamma }^{\prime }}%
\limits^{0}{}_{k}=w_{k}{}^{p}\left( {x}\right) \cdot L\left( {x}\right) %
\mathop {\gamma }\limits^{0}{}_{p}L^{-1}\left( {x}\right) =\mathop {\gamma }%
\limits^{0}{}_{k}\;.}
\end{array}
}\right.  \label{eq12}
\end{equation}
The matrix subscripts in $\mathop {\gamma} \limits^{0} {}_{k} $ respond to
the Lorentz transformations, however, taking into account that the vector
frame-of-reference subscript also responds to the Lorentz transformations,
it turns out that the resultant effect of the Lorentz transformations on $%
\mathop {\gamma} \limits^{0} {}_{k} $ is zero.

As $\gamma _{k} \left( {x} \right)$ and $\mathop {\gamma} \limits^{0} {}_{k} $
satisfy relations (\ref{eq9}), (\ref{eq11}) with one and the same metric
tensor $g_{mn} $ in the right-hand side, then, by the Pauli theorem, these
values should be related as
\begin{equation}  \label{eq13}
\gamma _{k} \left( {x} \right) = R\left( {x} \right)\mathop {\gamma }%
\limits^{0} {}_{k} R^{ - 1}\left( {x} \right),
\end{equation}
Here $R\left( {x} \right)$ is the field of the nonsingular matrix, which we
refer to as frame-of-reference matrix (the meaning of the name will become
clear right now). From (\ref{eq13}) and (\ref{eq8}) it follows that
\begin{equation}  \label{eq14}
\gamma _{\alpha} \left( {x} \right) = H_{\alpha} ^{k} \left( {x} \right)
\cdot R\left( {x} \right)\mathop {\gamma} \limits^{0} {}_{k} R^{ - 1}\left( {%
x} \right).
\end{equation}

The frame-of-reference matrix has a specific feature: its matrix subscripts
behave differently at invariant $T\left( {x} \right)$ transformations and at 
$L\left( {x} \right)$ transformations. From (\ref{eq12}), (\ref{eq13}) it
follows that at combination of the transformations
\begin{equation}  \label{eq15}
R\left( {x} \right) \to {R}^{\prime}\left( {x} \right) = T\left( {x}
\right)R\left( {x} \right)L^{ - 1}\left( {x} \right).
\end{equation}

If $T\left( {x} \right)$ transformations are considered as the world
transformations and $L\left( {x} \right)$ transformations as the local, then
it can be stated that in matrix $R\left( {x} \right)$ one subscript is of
world nature and the other of the local. The situation is similar to the one
which takes place in frame-of-reference vectors $H_{\alpha} ^{k} \left( {x}
\right)$, in which one subscript is associated with the world coordinates
and the second with the local as well. It is by virtue of this analogy that
matrix $R\left( {x} \right)$ is called frame-of-reference matrix. By the
way, in $\mathop {\gamma} \limits^{0} {}_{k} $ both vector and matrix
subscripts are local, that is why $\mathop {\gamma} \limits^{0} {}_{k} $ are
called doubly local.

Above we noted that availability of the discrete transposition operation
allowed construct nontrivial fields of matrix scalars, vectors, and other
tensors from DM. However, there is a class of DM, for which all nontrivial
scalar, vector and other fields can be made trivial, that is can be
converted to coordinate independent constants. A DM system of this type is
the ordinary Majorana system which will be used later on.

Along with the world invariant $T\left( {x} \right)$ transformations, local
transformations with similar features can be introduced. Denote the local
orthogonal transformations as $O\left( {x} \right)$ transformations.

Hereafter for product $R^{ - 1}\left( {x} \right) \cdot \Phi \left( {\gamma }
\right)$, where $\Phi \left( {\gamma} \right)$ is some scalar function of DM 
$\gamma _{\alpha} \left( {x} \right)$; $- \gamma _{\alpha} \left( {x}
\right);\;\gamma _{\alpha} ^{T} \left( {x} \right);\; - \gamma _{\alpha
}^{T} \left( {x} \right)$, we use notation $Z\left( {x} \right)$,
\begin{equation}  \label{eq16}
Z\left( {x} \right) = R^{ - 1}\left( {x} \right) \cdot \Phi \left( {\gamma}
\right).
\end{equation}
Object $Z\left( {x} \right)$ is called as bispinor matrix. In contrast to $%
R^{ - 1}$, matrix $Z$ can have zero determinant\footnote{%
The condition of matrix $Z$ equality to the nonsingular frame-of-reference
matrix discussed in ref. [3] is, strictly speaking, an additional
hypothesis. Our following consideration is not related to the hypothesis and
includes the case, where the rank of $Z$ is less than 4.} . At invariant $%
T\left( {x} \right)$ transformations and at the $L\left( {x} \right)$
transformations the bispinor matrix $Z\left( {x} \right)$ is transformed, as
it follows from (\ref{eq16}), in the same manner as matrix $R^{ - 1}$, that
is
\begin{equation}  \label{eq17}
Z\left( {x} \right) \to Z\left( {x} \right)R\left( {x} \right) = L\left( {x}
\right)Z\left( {x} \right)T^{ - 1}\left( {x} \right).
\end{equation}

From any DM system it is possible to construct a complete system of matrices 
$4 \times 4$, composed of 16 matrices. 

Hereafter of concern to us is implementation of the complete matrix systems
using the doubly local DM systems. The systems convenient for our purposes
are those composed of 10 symmetric and 6 antisymmetric matrices; these
systems appear in Table 1.

Table 1. Complete matrix systems

\begin{center}
\begin{tabular}{|c|c|c|}
\hline
& Symmetric matrices & Antisymmetric matrices \\ \hline
System 1 & $\mathop {\gamma} \limits^{0} {}_{k} \mathop {D}\limits^{0} {}^{
- 1};\;\mathop {S}\limits^{0} {}_{mn} \mathop {D}\limits^{0} {}^{ - 1} $ & $%
\mathop {D}\limits^{0} {}^{ - 1};\;\mathop {\gamma} \limits^{0} {}_{5} %
\mathop {\gamma} \limits^{0} {}_{k} \mathop {D}\limits^{0} {}^{ - 1};\;%
\mathop {\gamma} \limits^{0} {}_{5} \mathop {D}\limits^{0} {}^{ - 1} $ \\ 
\hline
System 2 & $\mathop {\gamma} \limits^{0} {}_{5} \mathop {\gamma} \limits^{0}
{}_{k} \mathop {C}\limits^{0} {}^{ - 1};\;\mathop {S}\limits^{0} {}_{mn} %
\mathop {C}\limits^{0} {}^{ - 1} $ & $\mathop {C}\limits^{0} {}^{ - 1};\;%
\mathop {\gamma} \limits^{0} {}_{k} \mathop {C}\limits^{0} {}^{ - 1};\;%
\mathop {\gamma} \limits^{0} {}_{5} \mathop {C}\limits^{0} {}^{ - 1} $ \\ 
\hline
\end{tabular}
\end{center}

If the field of vector $j^{\alpha} \left( {x} \right)$and the field of
antisymmetric tensor $H^{\alpha \beta} \left( {x} \right)$ are given, then
system 1 can be used to construct the scalar symmetric matrix,
\begin{equation}
Y\left( {x}\right) =\left( {{\textstyle{\frac{{1}}{{4}}}}j^{k}\left( {x}%
\right) }\right) \cdot \mathop {\gamma }\limits^{0}{}_{k}\mathop {D}%
\limits^{0}{}^{-1}+\left( {-{\textstyle{\frac{{1}}{{8}}}}H^{mn}\left( {x}%
\right) }\right) \cdot \mathop {S}\limits^{0}{}_{mn}\mathop {D}%
\limits^{0}{}^{-1}\quad .  \label{eq18}
\end{equation}

\noindent
at each point. Here 
\begin{equation}
\label{eq19}
j^{k}\left( {x} \right) = H_{\alpha} ^{k} \left( {x} \right)j^{\alpha 
}\left( {x} \right),\quad H^{mn}\left( {x} \right) = H_{\alpha} ^{m} \left( 
{x} \right)H_{\beta} ^{n} \left( {x} \right)H^{\alpha \beta} \left( {x} 
\right).
\end{equation}
Matrix fields of type (\ref{eq17}) are characterized with a certain type of their 
subscripts. Thus, in case (\ref{eq17}) field $Y\left( {x} \right)$ has two local 
subscripts. This allows, if necessary, changing components $j^{k}\left( {x} 
\right),\quad H^{mn}\left( {x} \right)$ in type (\ref{eq19}) expansions through 
Lorentz rotations of local frame-of-reference. 

When using $Y\left( {x} \right)$ field types, the invariance condition 
should be ensured: the subscript types in the left-hand and right-hand sides 
of the relations should coincide. If this condition is met, covariance and 
invariance with respect to $w^{m}_{n} \left( {x} \right)$ and $T\left( {x} 
\right)$ and $O\left( {x} \right)$ transformations is preserved in (\ref{eq17}) type 
expressions, despite using the doubly local DM and Roman tensor subscripts 
in these expressions. 

Matrix $Z$ can be represented as a direct sum of 4 bispinors using 
projectors $P_{\eta \lambda}  \;\left( {\eta ,\lambda = \pm}  \right)$ 
constructed from doubly local DM and satisfying the conditions of 
completeness and orthonormality:
\begin{equation}
\label{eq20}
\left. {\begin{array}{l}
 {P_{ + +}  + P_{ + -}  + P_{ - +}  + P_{ - -}  = E,} \\ 
 {P_{\eta \lambda}  \cdot P_{{\eta} '{\lambda} '} = \delta _{\eta {\eta} '} 
\delta _{\lambda {\lambda} '} P_{\eta \lambda}  \quad .} \\ 
 \end{array}}  \right\}
\end{equation}

The representation of matrix $Z$ is:
\begin{equation}
\label{eq21}
Z = ZP_{ + +}  + ZP_{ + -}  + ZP_{ - +}  + ZP_{ - -}  .
\end{equation}
Separate addends $\Psi _{\eta \lambda}  \equiv ZP_{\eta \lambda}  $ in the 
right-hand side of (\ref{eq21}) have 4 parameters and are transformed at the Lorentz 
transformations of local frame-of-reference according to law

\begin{equation}
\label{eq22}
{\Psi} '_{\eta \lambda}  \left( {x} \right) = L\left( {x} \right)\Psi _{\eta 
\lambda}  \left( {x} \right).
\end{equation}

 $\Psi _{\eta \lambda}  $ is equivalent to ordinary 4-component column bispinor.

Write the bispinor matrix as a so-called polar expansion
\begin{equation}
\label{eq23}
Z\left( {x} \right) = H\left( {x} \right) \cdot U^{ - 1}\left( {x} \right),
\end{equation}
Here $H\left( {x} \right)$ is a symmetric and nonnegatively definite matrix 
and $U^{ - 1}\left( {x} \right)$ is an orthogonal one. The possibility that 
any square real matrix can be represented as (\ref{eq23}) follows from the classic 
theory of matrices. Here multiplier $H\left( {x} \right)$ is uniquely 
defined in the polar expansion. We call matrix $H\left( {x} \right)$ as 
amplitude and $U^{ - 1}\left( {x} \right)$ as phase.

Consider behavior of each multiplier in (\ref{eq23}) at the Lorentz transformations 
of local frame-of-reference. At such transformations, according to Table 1, 
the bispinor matrix is transformed by law
\begin{equation}
\label{eq24}
Z\left( {x} \right) \to {Z}'\left( {x} \right) = L\left( {x} \right)Z\left( 
{x} \right).
\end{equation}
Upon the Lorentz transformation of the local frame-of-reference the bispinor 
matrix can again be represented in a form similar to (\ref{eq23}),
\begin{equation}
\label{eq25}
{Z}'\left( {x} \right) = {H}'\left( {x} \right) \cdot {U'}^{ - 1}\left( {x} 
\right).
\end{equation}
${H}'\left( {x} \right)$ and ${U'}^{ - 1}\left( {x} \right)$, as it follows 
from (\ref{eq23}), (\ref{eq24}), (\ref{eq25}), should satisfy relation:
\begin{equation}
\label{eq26}
{H}'\left( {x} \right) \cdot {U}'^{ - 1}\left( {x} \right) = L\left( {x} 
\right) \cdot H\left( {x} \right) \cdot U^{ - 1}\left( {x} \right).
\end{equation}
However, relation (\ref{eq26}) by no means dictates any definite transformation 
rules for each of the multipliers in the polar expansion. The rules can be 
specified only at special form of matrices $L\left( {x} \right)$: when the 
matrices describe spatial rotations and, hence, are orthogonal, that is when 
\begin{equation}
\label{eq27}
L\left( {x} \right) = R\left( {x} \right),\quad R^{T}\left( {x} \right) = 
R^{ - 1}\left( {x} \right).
\end{equation}
In the case of (\ref{eq27}) the multipliers are transformed as follows:
\begin{equation}
\label{eq28}
\left. {\begin{array}{l}
 {H\left( {x} \right) \to {H}'\left( {x} \right) = R\left( {x} 
\right)H\left( {x} \right)R^{T}\left( {x} \right),\;} \\ 
 {U^{ - 1}\left( {x} \right) \to {U'}^{ - 1}\left( {x} \right) = R\left( {x} 
\right)U^{ - 1}\left( {x} \right)\;.} \\ 
 \end{array}}  \right\}
\end{equation}
The fact that in the general case there is no definite law of transformation 
for each of the multipliers in the polar expansion leads to an important 
conclusion, which is formulated below. Keeping in mind that the amplitude is 
a symmetric matrix, we can expand it by a complete system of symmetric 
matrices, for example, by system 1 from Table 1. 
\begin{equation}
\label{eq29}
H\left( {x} \right) = - v_{0} \cdot \mathop {\gamma} \limits^{0}{}_{0} 
\mathop {D}\limits^{0} {}^{ - 1} + v_{b} \cdot \mathop {\gamma} \limits^{0} 
{}_{b} \mathop {D}\limits^{0} {}^{ - 1} + {w_{ab}}\cdot\mathop S\limits^{0} 
{}_{ab} \mathop {D}\limits^{0} {}^{ - 1} -2 {w_{0b}}\cdot\mathop S\limits^{0} 
{}_{0b} \mathop {D}\limits^{0} {}^{ - 1} \quad .
\end{equation}
The coefficients in the expansion are found in a standard manner. However, 
expansion coefficients $\left( {v_{0} ,v_{b}}  \right),\;\left( {w_{ab} 
,w_{0b}}  \right)$ are not components of the local vector and the 
antisymmetric tensor, respectively. If they had been such, expression (\ref{eq24}) 
would have been invariant and transformed by rule $H\left( {x} \right) \to 
{H}'\left( {x} \right) = L\left( {x} \right)H\left( {x} \right)L^{T}\left( 
{x} \right)$ at the Lorentz transformations of the local frame-of-reference. 
But the rule, together with the rule of transformation of bispinor matrix 
(\ref{eq24}), lead to the fact that the phase multiplier has to be transformed by 
rule $U^{ - 1}\left( {x} \right) \to {U'}^{ - 1}\left( {x} \right) = L^{T}{}^{- 1}
\left( {x} \right)U^{ - 1}\left( {x} \right)$. The last rule leads to a 
contradiction: on the Lorentz transformation the phase multiplier is no 
longer orthogonal. It is this contradiction that the above statement follows 
from: coefficients $\left( {v_{0} ,v_{b}}  \right),\;\left( {w_{ab} ,w_{0b} 
} \right)$ do not generate the local vector and the antisymmetric tensor, 
respectively. 

Note that the situation in expansion (\ref{eq29}) of amplitude matrix $H\left( {x} 
\right)$ basically differs from expansion (\ref{eq17}) for scalar symmetric matrix 
field $Y\left( {x} \right)$. One difference is that in the case of $Y\left( 
{x} \right)$the vector field and the antisymmetric tensor field were given a 
priori, while in the case of $H\left( {x} \right)$ there was no similar a 
priori requirement. The other difference is that in the case of $H\left( {x} 
\right)$ it is additionally required that product $H\left( {x} \right) \cdot 
U^{ - 1}\left( {x} \right)$ have the same subscript type as the bispinor 
matrix.

3. Formulation of the principal proposition 

Assume that five real tensor values presented in Table 2 are given at some 
Riemann space point.

Table 2. List of tensors

\begin{center}
\begin{tabular}{|c|c|}
\hline
Tensor value & Notation \\ \hline
Scalar & $m$ \\ \hline
Vector & $j^{\alpha} $ \\ \hline
Pseudo-vector & $s_{\alpha} $ \\ \hline
Anti-symmetric tensor & $H_{\alpha \beta} $ \\ \hline
Pseudo-scalar & $n $ \\ \hline
\end{tabular}
\end{center}

Assume that an arbitrary, but fixed implementation of the Dirac symbols as 
complex matrices 4$ \times $4 is taken. Let $D$ be a nonsingular matrix 
connecting two local Dirac matrix systems, $\gamma _{k} $ and $ - \gamma 
_{k}^{ +}  $,
\begin{equation}
\label{eq30}
D\gamma _{k} D^{ - 1} = - \gamma _{k}^{ +}  .
\end{equation}
Matrix $D$ is determined by relation (\ref{eq1}) with an accuracy of multiplication 
by an arbitrary complex number. Using this freedom, it is always possible to 
make that matrices $D,\;D^{ - 1}$ be anti-Hermitean.$D^{ - 1}$and Dirac 
matrices can be used to construct the following complete system of matrices 
4$ \times $4 composed solely of Hermitean matrices:
\begin{equation}
\label{eq31}
 - iD^{ - 1};\quad \gamma _{\alpha}  D^{ - 1};\quad - i\gamma _{5} \gamma 
_{\alpha}  D^{ - 1};\quad - S_{\alpha \beta}  D^{ - 1};\quad i\gamma _{5} 
D^{ - 1}
\end{equation}
Construct matrix $M$ with the following algorithm using the given tensors 
and Hermitean matrix system (\ref{eq31}):
\begin{equation}
\label{eq32}
M ={{1}\over{4}}\left( - iD^{ - 1} m + \gamma 
_{\alpha}  D^{ - 1} j^{\alpha}  
- i\gamma _{5} \gamma _{\alpha}  D^{ - 1} s^{\alpha} 
- S_{\alpha \beta}  D^{-1} 
\left( {{\textstyle{{1} \over {2}}}H^{\alpha \beta} } \right) + 
i\gamma _{5} D^{-1}  n \right) 
\end{equation}
Assume that the matrix $M$ is non-negatively definite, that is all of its 
four eigenvalues $\left( {\lambda _{1} ,\lambda _{2} ,\lambda _{3} ,\lambda 
_{4}}  \right)$are nonnegative. Denote the rank of the matrix $M$ as $r$; 
clear that $r$ can take values from 0 to 4, with the rank of the matrix $M$ 
being able to become zero only when $M = 0$.

If the nonnegativity condition is met, then valid are 

\underline {Propositions}

1. Let $H$ be Hermitean matrix in binary expansion $M = HH^{ +} $. There are 
no more than $2^{r}$ unitarily nonequivalent matrices $H$, each of which is 
correspondent with one and the same set of tensors listed in Table 2. By the 
unitary nonequivalence is meant that different Hermitean multipliers $H_{1} 
,\;H_{2} $ in the binary expansion can not be related as $H_{1} = U_{12} 
H_{2} U_{12}^{ +}  $, where $U_{12} $ is unitary. 

2. Matrix $Z$ coincides with arithmetic root of matrix $M$ with an accuracy 
of the unitary multiplier $U$ on the right, that is $Z = hU$, where $h$ is 
the matrix among matrices $H$, which is nonnegative.

3. The tensors listed in Table 2 relate to each of matrices $Z$ as
\begin{equation}
\label{eq33}
\begin{array}{l}
 m \equiv iSp\left( {Z^{ +} DZ} \right)\;,\quad j^{\alpha}  \equiv Sp\left( 
{Z^{ +} D\gamma ^{\alpha} Z} \right)\;,\quad s_{\alpha}  \equiv iSp\left( 
{Z^{ +} D\gamma _{5} \gamma _{\alpha}  Z} \right)\;,\quad \\ 
 H_{\alpha \beta}  \equiv Sp\left( {Z^{ +} DS_{\alpha \beta}  Z} 
\right)\;,\quad n \equiv iSp\left( {Z^{ +} D\gamma _{5} Z} \right) \\ 
 \end{array}
\end{equation}
4. Solution of the problem of finding matrix $Z$ with using Dirac matrices 
for basis

Write the expansion for $Z$ through Dirac matrices:
\begin{equation}
\label{eq34}
Z = a \cdot E + iA_{0} \cdot \gamma _{0} + A_{k} \cdot \gamma _{k} + iB_{0} 
\cdot \gamma _{5} \gamma _{0} + B_{k} \cdot \gamma _{5} \gamma _{k} + ib 
\cdot \gamma _{5} + C_{k} \cdot \gamma _{0} \gamma _{k} + ih_{k} \cdot 
\gamma _{5} \gamma _{0} \gamma _{k} \quad .
\end{equation}
As we assume that the matrix $Z$ is Hermitean, it follows that in expansion 
(\ref{eq34}) the coefficients of the Hermitean matrices are real and those of the 
anti-Hermitean are imaginary. 

Multiply matrix $Z$ by matrix $Z^{ +} $. 
\begin{equation}
\label{eq35}
\begin{array}{l}
 ZZ^{ +}  = \\ 
 = E \cdot \left\{ {a^{2} + A^{2} + B^{2} + C^{2} + b^{2} + A_{0}^{2} + 
B_{0}^{2} + h^{2}} \right\} + \\ 
 + 2\gamma _{k} \cdot \left\{ {aA_{k} + \varepsilon _{kab} B_{a} C_{b} + 
B_{0} h_{k}}  \right\} + 2\gamma _{5} \gamma _{k} \cdot \left\{ {aB_{k} - 
\varepsilon _{kab} A_{a} C_{b} - A_{0} h_{k}}  \right\} + \\ 
 + 2\gamma _{0} \gamma _{k} \cdot \left\{ {aC_{k} + \varepsilon _{kab} A_{a} 
B_{b} + bh_{k}}  \right\} + 2i\gamma _{5} \cdot \left\{ {ab + \left( {Ch} 
\right)} \right\} + 2i\gamma _{0} \cdot \left\{ {aA_{0} - \left( {Bh} 
\right)} \right\} + \\ 
 + 2i\gamma _{5} \gamma _{0} \cdot \left\{ {aB_{0} + \left( {Ah} \right)} 
\right\} + 2i\gamma _{5} \gamma _{0} \gamma _{k} \cdot \left\{ {ah_{k} + 
A_{k} B_{0} - A_{0} B_{k} + bC_{k}}  \right\} \\ 
 \end{array}
\end{equation}
Set obtained expression (\ref{eq34}) equal to expression (\ref{eq31}). 
\begin{equation}\label{eq36}
\left. {\begin{array}{l}
 {a^{2} + A^{2} + B^{2} + C^{2} + b^{2} + A_{0}^{2} + B_{0}^{2} + h^{2} = 
{\textstyle{{1} \over {4}}}j^{0}} \\ 
 {aA_{k} + \varepsilon _{kab} B_{a} C_{b} + B_{0} h_{k} = - {\textstyle{{1} 
\over {8}}}H_{0k}}  \\ 
 {aB_{k} - \varepsilon _{kab} A_{a} C_{b} - A_{0} h_{k} = {\textstyle{{1} 
\over {16}}}\varepsilon _{kpq} H_{pq}}  \\ 
 {aC_{k} + \varepsilon _{kab} A_{a} B_{b} + bh_{k} = {\textstyle{{1} \over 
{8}}}j_{k} ;\quad ab + \left( {Ch} \right) = {\textstyle{{1} \over 
{8}}}s_{0} ;\quad aA_{0} - \left( {Bh} \right) = {\textstyle{{1} \over 
{8}}}m} \\ 
 {aB_{0} + \left( {Ah} \right) = - {\textstyle{{1} \over {8}}}n;\quad ah_{k} 
+ A_{k} B_{0} - A_{0} B_{k} + bC_{k} = - {\textstyle{{1} \over {8}}}s_{k}}  
\\ 
 \end{array}}  \right\}
\end{equation}
If it is possible to solve system (\ref{eq36}), then the answer to the question 
formulated in the title of this section will be given. The solution 
procedure consists in expression of the values appearing in the left-hand 
sides of the equations, 
\begin{equation}
\label{eq37}
a,\quad b,\quad A_{0} ,\quad A_{k} ,\quad B_{0} ,\quad B_{k} ,\quad C_{k} 
,\quad h_{k} ,
\end{equation}
through those appearing in the right-hand sides of the equations, 
\begin{equation}
\label{eq38}
j_{0} ,\quad H_{0k} ,\quad H_{pq} ,\quad j_{k} ,\quad s_{0} ,\quad m,\quad 
n,\quad s_{k} \quad .
\end{equation}
System (\ref{eq36}) becomes somewhat reduced in the number of unknowns, if we are 
manipulating over the real number field. In doing so only the following 
remain from among values (\ref{eq37}), (\ref{eq38}):
\begin{equation}
\label{eq39}
a,\quad A_{k} ,\quad B_{k} ,\quad C_{k} \quad ;
\quad
j_{0} ,\quad F_{0k} ,\quad F_{pq} ,\quad j_{k} \quad .
\end{equation}
If the general normalization of desired values to $\sqrt {{\textstyle{{1} 
\over {4}}}j^{0}} $ and given values to ${\textstyle{{1} \over {4}}}j^{0}$ 
is introduced, then, taking
\[
\vec {x} \equiv \frac{{\vec {A}}}{{\sqrt {{\textstyle{{1} \over {4}}}j^{0}} 
}};\vec {y} \equiv \frac{{\vec {B}}}{{\sqrt {{\textstyle{{1} \over 
{4}}}j^{0}}} };\vec {z} \equiv \frac{{\vec {C}}}{{\sqrt {{\textstyle{{1} 
\over {4}}}j^{0}}} };a_{k} \equiv - \frac{{1}}{{2}}\frac{{H_{0k} 
}}{{j^{0}}};b_{k} \equiv \frac{{1}}{{4}}\varepsilon _{kpq} \frac{{H_{pq} 
}}{{j^{0}}};c_{k} \equiv \frac{{1}}{{2}}\frac{{j_{k}} }{{j^{0}}}
\]
system (\ref{eq36}) becomes:
\begin{equation}
\label{eq40}
a^{2} + \vec {x}^{2} + \vec {y}^{2} + \vec {z}^{2} = 1;a \cdot \vec {x} + 
\left[ {\vec {y};\vec {z}} \right] = \vec {a};a \cdot \vec {y} + \left[ 
{\vec {z};\vec {x}} \right] = \vec {b};a \cdot \vec {z} + \left[ {\vec 
{x};\vec {y}} \right] = \vec {c}.
\end{equation}
Now it is possible to construct solutions to the last nine equations of 
system (\ref{eq40}) explicitly.

A solution to the characteristic equation for matrix$M$, defined by formula 
(\ref{eq32}) is 
\begin{equation}
\label{eq41}
\left. {\begin{array}{l}
 {\lambda _{1} = j + \sqrt {u^{2} + v^{2} + 2w} ;\quad \lambda _{2} = j + 
\sqrt {u^{2} + v^{2} - 2w}}  \\ 
 {\lambda _{3} = j - \sqrt {u^{2} + v^{2} - 2w} ;\quad \lambda _{4} = j - 
\sqrt {u^{2} + v^{2} + 2w}}  \\ 
 \end{array}}  \right\} \quad .
\end{equation}
Here the following notations are used:
\begin{equation}
\label{eq42}
\left. {\begin{array}{l}
 {j \equiv \sqrt {\left( { - g_{\mu \nu}  j^{\mu} j^{\nu} } \right)} \ge 
0\;,\quad e^{\alpha}  \equiv {{j^{\alpha} } \mathord{\left/ {\vphantom 
{{j^{\alpha} } {\sqrt {\left( { - g_{\mu \nu}  j^{\mu} j^{\nu} } \right)} 
}}} \right. \kern-\nulldelimiterspace} {\sqrt {\left( { - g_{\mu \nu}  
j^{\mu} j^{\nu} } \right)}} }\;,} \\ 
 {u_{\alpha}  \equiv H_{\alpha \nu}  e^{\nu} ,\quad u^{2} \equiv g^{\alpha 
\beta} u_{\alpha}  u_{\beta}  \;,} \\ 
 {v_{\alpha}  \equiv {\textstyle{{1} \over {2}}}E_{\alpha \mu \nu \lambda}  
H^{\mu \nu} e^{\lambda} \;,\quad v^{2} \equiv g^{\alpha \beta} v_{\alpha}  
v_{\beta}  \;,} \\ 
 {w_{\alpha}  \equiv \left[ {\delta _{\alpha} ^{\beta}  + e_{\alpha}  
e^{\beta} } \right]H_{\beta \mu}  H^{\mu \nu} e_{\nu}  \;,\quad w \equiv 
\sqrt {g^{\alpha \beta} w_{\alpha}  w_{\beta} }  \quad .} \\ 
 \end{array}}  \right\} \quad .
\end{equation}
From formulas (\ref{eq42}) it follows, that vectors $u^{\alpha} ,\;v^{\alpha 
},\;w^{\alpha} $ are orthogonal to vector $j^{\alpha} $, in addition, 
vectors $u^{\alpha} ,\;v^{\alpha} $ are orthogonal to vector$\;w^{\alpha} $. 
The least eigenvalue is $\lambda _{4} $. For the matrix $M$ to be 
non-negative, the following condition has to be met:
\begin{equation}
\label{eq43}
j \ge \sqrt {u^{2} + v^{2} + 2w} \quad .
\end{equation}
Inequality (\ref{eq43}) is the only condition for solvability of our problem.

Then matrix $Z$ was constructed explicitly with the diagonalizing matrix 
$V$. The result was validated with a computer program of symbol 
computations. 

The mapping of the world tensors on the amplitude part in the polar 
expansion of the bispinor matrix found at a choice of local 
frame-of-reference in the tangent spaces of the Riemannian space is retained 
at the following transformations:
world coordinate transformations;
invariant $T\left( {x} \right)$ transformations;
invariant $O\left( {x} \right)$ transformations.

The first, the second and third properties are evident, since none of the 
values appearing in the mapping relation explicitly contains either Greek or 
world matrix subscripts. 

As for the mapping invariance with respect to the Lorentz transformations of 
local frame-of-reference, this has to be elucidated, since the amplitude 
multiplier has no definite type of matrix subscripts. The point is that 
amplitude multiplier $H\left( {x} \right)$ appears in a quadratic 
combination in the mapping relation $M\left( {x} \right) = H\left( {x} 
\right)H^{T}\left( {x} \right)$. The combination has the same matrix 
subscript type as matrix $M$, since $HH^{T} = ZZ^{T}$. Therefore, if the 
mapping has been found at one frame-of-reference set, it will be also 
retained at any other, except for the following: at a transition to a new 
frame-of-reference field matrix $H'\left( {x} \right)$ has to be an 
amplitude multiplier for the new bispinor matrix, $Z'\left( {x} \right) = 
H'\left( {x} \right) \cdot U'^{ - 1}\left( {x} \right)$, which relates to 
the old one as $Z'\left( {x} \right) = L\left( {x} \right)Z\left( {x} 
\right)$. 

5. Comments 

When describing half-integer spin particles, the physical values observed 
can be only compared with scalar, vector or other tensor items which are 
quadratic in bispinor components. There are no experiments with such 
particles whose results would be expressible through items constructed from 
odd degrees of the bispinor components, i.e. the components themselves are 
not physically observable items. Hence, the total amount of information 
about the particle spin structure is contained in the space-time geometry 
structure. Therefore, a complete description of half-integer intrinsic 
moment particles does not require any special items, like bispinors, but can 
be uniquely expressed solely in terms of space-time tensor values, i.e. in 
terms of Riemannian variety invariants.

The mapping between the Riemannian variety tensor items and matrix 
$Z$ composed of four bispinors constructed in this paper solves the problem. 
The uniqueness in the proposed construction is achieved thanks to the fact 
that at $Z$ matrix generation by the tensor invariants the entire 
arbitrariness is localized in the gauge transformation unitary matrix 
responsible for the internal degrees of freedom of half-integer spin 
particles. Hence we arrive at a very uncommon physical corollary: the 
existence of the spin structure of particles as of a purely geometric item 
of the Riemannian variety is only possible, given other internal degrees of 
freedom in the particles, with the gauge symmetry group describing these 
degrees of freedom being uniquely determined by the Riemannian variety 
dimension $``n"$ in accordance with the chain $n \leftrightarrow N \times N$ 
- dimension $N$ (=dimensions of matrices $Z$) of the Dirac matrices in space 
$ \leftrightarrow $ dimension of the unitary group of matrices $U$ acting on 
the matrix $Z$ on the right $ \leftrightarrow $ gauge group.

The case of four-dimensional matrices $Z_{4R} $ over the real number field 
considered in the paper corresponds to nonzero spin and zero electric charge 
particles. In particular, neutron and neutrino are such particles. Our 
conclusion agrees with presently known experimental data: all 
spin-containing neutral particles are members of multiplets in internal 
symmetry group representations. 

Note that the gauge group appearing in the case of $Z_{4R} $ is group 
$O\left( {4} \right)$ which is direct product $SO\left( {3} \right) \times 
SU\left( {2} \right)$(+ all non-intrinsic automorphisms in this product). 
Physically, this is a direct hint about possible correspondence with the 
gauge group of electroweak interactions. 

The second most important corollary of the spinor structure geometrization 
method proposed in the paper is that the $Z$ matrix field dynamics 
completely depends on tensor fields, which eventually reduces to dynamic 
equations for the Riemannian variety. Naturally, the constructive derivation 
of the general dynamic equations for $Z$ matrix from equations of the 
general relativity theory (GRT) will require solution of several very 
difficult problems pertaining to the GRT equations themselves. These are 
primarily: the problem of representation of Riemannian variety global 
properties which are solutions to dynamic equations. The second problem 
group pertains to development of algebraic methods for integration of 
partial differential equations which allow expression of solutions in terms 
of Riemannian variety tensor algebraic invariants. Finally, the third 
problem is construction of the energy-momentum tensor, i.e. the right-hand 
side in the GRT equations, at which the solutions to the equations provide 
the tensor invariants defined on the Riemannian variety which lead to 
non-negatively definite matrix $M$ (see formula (\ref{eq32})). It turns out that our 
spinor geometrization method automatically leads to formulation of 
conditions for the Riemannian variety global structure and then, eventually, 
for the energy-momentum tensor property. These conditions result from 
studying the covariant expressions for the tensor invariants comprising the 
condition of $M$ matrix non-negativity (see formulas (\ref{eq42}) and (\ref{eq43})), namely: 
a) existence of globally-definite time-like vector $j^{\alpha} $, i.e. 3+1 
structure of the Riemannian variety; b) existence of global partition 2+1 on 
3D hyper-surface (i.e. two vectors, $u_{\alpha}  $ and $v_{\alpha}  $, 
generate a 2D surface orthogonal to vector $w_{\alpha}  $, see formula 
(\ref{eq42})).

Summing up items a) and b), the result can be expressed as a single 
condition: for the geometrized spin structure to exist, the Riemannian 
variety has to admit the 2+2 global structure (naturally, we can not argue 
here that it is this that the whole set of sufficient conditions for 
$M$matrix non-negativity reduces to). It should be noted that this 
conclusion completely agrees with J. Wheeler's guess about potential 
properties of the Riemannian varieties necessary for adequate description of 
$1/2$-spin particles. 

The work was supported by the International Science and Technology Center (Project \#KR-154).

\vskip5mm

\noindent
[1] M.V.Gorbatenko, A.V.Pushkin. \textit{VANT. Ser. Teor. Prikl. Fizika.} 
\textbf{3}, 3 (1999).

\noindent
[2] M.V.Gorbatenko, A.V.Pushkin. \textit{Ibid:} 
\textbf{3}, 19 (1999).

\noindent
[3] M.V.Gorbatenko, A.V.Pushkin. \textit{Ibid:} 
\textbf{1}(\textbf{1}), 49 (1984). M.V.Gorbatenko. \textit{Theoretical and 
Mathematical Physics}, \textbf{103}, No. 1 (1995) 374.

\end{document}